\documentstyle[prb,multicol,aps,epsf]{revtex}
\newcommand{\be}{\begin{equation}}
\newcommand{\ee}{\end{equation}}
\newcommand{\bea}{\begin{eqnarray}}
\newcommand{\eea}{\end{eqnarray}}

\newcommand{\p}{\partial}

\newcommand{\re}{{\rm e}}
\newcommand{\rd}{{\rm d}}
\newcommand \varep {\varepsilon}
\begin {document}
\bibliographystyle {plain}

\title
{Non-existence of strong coupling two-channel Kondo fixed point
for microscopic models of tunneling centers.}

\author{I. L. Aleiner $^{(a,b)}$ and D. Controzzi$^{(c)}$   
}

\address{$^{(a)}$ Department of Physics and Astronomy, SUNY at Stony Brook, 
NY 11794, USA 
\\
$^{(b)}$ Physics Department, Lancaster University, 
LA1 4YB, Lancaster, UK 
\\
$^{(c)}$Department of Physics, Princeton University, 
Princeton NJ 08544, USA
}

\maketitle

\begin{abstract}
\par
We consider the problem of an heavy particle in a double well potential (DWP) 
interacting with an electron bath. Under general assumptions, we map 
the problem to a 
three-color 
logarithmic gas model, where the size of
the core of the 
charged particles 
is proportional to the tunneling time, $\tau_{tun}$,
of the heavy particle between the two
wells. For times larger then $\tau_{tun}$ this model is equivalent to
the anisotropic two-channel Kondo (2CK) model in a transverse field.
This allows us to establish a  relationship between the 
microscopic parameters of DWP and the 2CK problem. We show that 
the strong coupling fixed point of the 2CK
model can never be reached for the DWP problem, 
in agreement with the results of 
Kagan and Prokof'ev, 
[Sov.Phys. JEPT, {\bf 69}, 836 (1989)]
and Aleiner {\em et al.}, [\prl {\bf 86}, 2629 (2001)].
\end{abstract}

\sloppy
\begin{multicols}{2}

\section{Introduction}
The possibility to observe the two channel Kondo effect in real physical
systems
has attracted a lot of interest since the pioneering work by Nozi\'eres and 
Blandin\cite{NB}, 
but at present
no experimental realization has been conclusively demonstrated. The 
difficulty lies in the fact that the non-Fermi liquid (NFL)
fixed point of the 
two channel Kondo model \cite{nfl} 
is unstable to various symmetry breakings 
\cite{NB,ALC} which turn
out to be important for various experimental situations. In particular, 
channel anisotropy is a relevant perturbation and the exact channel 
symmetry is required for the NFL
physics to be observed. In a conventional magnetic realization of
the Kondo effect this would require 
the exact degeneracy between atomic orbitals that cannot be obtained in any 
real system. In the search for 2CK effect,
systems which are less directly related to
the conventional Kondo physics were considered. 
It was suggested that 
non magnetic impurity tunneling between two 
sites and interacting with an electron bath 
could be modeled as
a two channel Kondo system in which the spin plays the role of the channel 
index \cite{StromOlson,Kondo76,VZ,KP1,KP2} 
(see also Ref.~\onlinecite{CoxZaw,exp.rev} for a review). 
Following this suggestion,
the NFL behaviour of the 
2CK fixed point  was used in Ref.~\onlinecite{kondo.exp}
 to interpret the low temperature
transport data for narrow  metallic constrictions.
Indeed, in the limit $k_F a\ll1$, where $k_F$ is the Fermi wave 
vector and $2a$ is the distance between the two minima of the DWP, 
only electrons with two spherical harmonics ($l=0$ and  $l=1, \;m=0$) strongly 
interact with the heavy particle. 
Usually the  mapping of the DWP into a two level system (TLS), which behaves 
like a localized spin, is achieved
by restricting the Hilbert space of the atom to the two lowest
energy states associated with the two minima. 
(As we will discuss in the 
following this approach is not well justified because it fails
to capture the connection between the coupling constants of the effective
theory.)
Then the orbital degrees of freedom play the
role of a pseudo-spin while the real 
electron spin represent the channel index, and,
in absence of magnetic field, the channel degeneracy was 
guaranteed by construction.
A disadvantage of this realization for the 2CK effect 
is that other relevant terms appear 
which are no longer forbidden by symmetry,
the most important being the spontaneous tunneling of the impurity between the
two minima. This  corresponds to magnetic field in the conventional Kondo 
problem and  then the  resulting Kondo model
has the form 
\be
H_{Kondo}=H^0_{k}
+H_I
+h S^x +\Delta_z S^z.
\label{kondo}
\ee
Here 
\be
H^0_{k}=\int \rd q \;\epsilon_{ q} \;
\psi^{\dagger}_{\alpha,i}
({ q})
\psi_{\alpha,i}({ q}) 
\ee
represents the propagation of free band electrons, 
$\alpha=1,2$ is the pseudo-spin index and  $i=1,2,...,k$ 
is the channel index. In the specific case $k=2$.
The sum over repeated indices is implied. The second term in (\ref{kondo})
describes the interaction between the localized impurity, $\vec S$,
and the electrons
\be
H_I=\frac{1}{\nu}\sum_{a=x,y,z}\lambda_a \psi^{\dagger}_{\alpha,i}({ r})
\sigma^a
_{\alpha \beta}
\psi_{\beta,i}({ r})S^a \Bigg|_{r=0},
\ee
where  $\sigma^a_{\alpha\beta}$ are the Pauli matrices in the
pseudo-spin space, the fermionic operators 
$\psi_{\alpha,i}({ r})$ are the counterparts of $\psi_{\alpha,i}({ k})$
in the coordinate representation, and  
$\nu$ is the density of electronic states at the Fermi energy per one channel, 
which is 
introduced here to make the coupling constants dimensionless.
There are two terms in the Hamiltonian (\ref{kondo}) that break the rotational
symmetry in the pseudo-spin space.
The term proportional to 
$\Delta_z$ is related to the original asymmetry of the DWP, 
and may be greatly enhanced by the disordered potential 
acting on electrons\cite{WAM}. However, it is still possible
to imagine, that the original DWP is symmetric, and all the
other impurities are remote from the DWP, so the value
of  $\Delta_z$ is negotiable and may be even set to zero in some 
particular cases.
Nevertheless, the effective magnetic field, $h$, is related to the 
tunneling of the impurity in the DWP, and so is the constant $\lambda_x$.
The relation between them requires microscopic consideration and one is not
allowed to neglect $h$ but keep $\lambda_x$ finite.

The model (\ref{kondo}) is known to scale to the strong coupling fixed
point \cite{NB,AL}, 
where it has a NFL behaviour\cite{nfl}. This regime is achieved if 
both temperature, $T$, and 
\be
\Delta=\sqrt{\Delta_z^2+h^2}
\ee
do not exceed the Kondo temperature, $T_K$.
Since the interaction in highly anisotropic, $\lambda_y$ can be chosen to be
equal to zero and 
$\lambda_x \ll \lambda_z$, 
the Kondo temperature is given by
(see for instance Ref.~\onlinecite{jnw})
\be
T_K=D (\lambda_x \lambda_z)^{1/2} \left ( \frac{\lambda_x}{2\lambda_z}
\right )^{1/2\lambda_z},
\label{tk}
\ee
where $D$ is a high energy cut-off.
In order to check whether the 2CK effect can be observed
in a DWP system one then 
has to  compare $T_k$ with $\Delta$. This problem has recently 
attracted a lot 
of attention  \cite{VZ,KP1,KP2,ZZ,FM2,zznvz,AAGS,ZZ2,AAG}. 
Some alarming results already appeared in 
\cite{KP2} and were confirmed in \cite{AAGS}.
The crucial point consists in determining the various coupling constants 
in (\ref{kondo}) and also high-energy cut-off $D$ in Eq.~(\ref{tk})
 starting from a general microscopic description of the 
problem. 

As first pointed out by Kagan and Prokof'ev \cite{KP1,KP2},
the TLS is not a good starting point since the high energy degrees of freedom 
of the impurity cannot be taken into account and they turn out to be essential
to establish the correct mapping to the Kondo problem \cite{KP1,KP2,AAGS}.
In particular it was shown by model calculation of
Ref.~\onlinecite{AAGS} that $D$ is of the order of the energy
distance to the third excited level of the atomic system which is several
orders of magnitude smaller than the Fermi energy. Together with
the restrictions to the coupling constant it leads to $h > T_K$ and
thus to the 
impossibility to reach the strong coupling limit. However,
a question
arises, whether those conclusions are model-dependent or there is a general
principle for wide class of microscopic models preventing the existence
of the strong coupling regime. 
 
In this paper we use a general non-perturbative approach  
to the problem that takes into account
all the states of the DWP and that, we believe,
provides a conclusive answer to the question
weather it is possible to observe the 2CK effect in the problem of
a tunneling impurity.  We start from a microscopic description of the
moving heavy particle interacting with electrons in a metal. The
potential in which the impurity moves is chosen to have  the most
general form with two minima (the situation of a potential that
presents three minima \cite{MF} is not considered here).  The
interaction of the heavy atom
with the electrons is chosen to be instantaneous density-density
interaction, 
which, we believe, contains all the essential physics; effects of retardation
due to the electron screening or the inter-band excitations are small
either as inverse band gap or plasma frequency and will not be taken
into account.

Using
a semi-classical approximation to describe the dynamics of the
impurity and the condition $k_Fa\ll 1$ we map the problem to a
one-dimensional logarithmically interacting gas  model (LGM). 
Since a similar mapping can be
also obtained for the Kondo model\cite{AY,FGN}, this allows us to
establish a general relationship between the microscopic parameters of
the DWP and the coupling constants of the Kondo model.  
Our results are in agreement with the
predictions of Refs.~\onlinecite{KP1,KP2,AAGS} and 
show that the mapping of the DWP to
the 2CK model can be done only for energies below $\hbar
\tau_{tun}^{-1}$, where $\tau_{tun}$ is the tunneling time of the heavy
particle between the two wells. It is the existence of this small
energy scale, together with the relationship between $h$ and $\lambda_x$,
that makes the Kondo temperature too small for the strong
coupling fixed point to be achieved in such systems: $T_K$
turns out to be always smaller than $\Delta$, which makes the strong
coupling limit non-accessible. 

The paper is organized as follow. In the next Section we describe the model and
we map it to a logarithmic gas model in Sec. III. In Section IV we 
obtain the relation between our microscopic model and the 2CK model and,
in the last Section, we  
summarize the results and draw some general conclusions. The problem of the
effect of the electron-hole asymmetry on our results is discussed in the 
Appendix.

{\begin{figure}[ht]
\epsfxsize=6cm
\centerline{\epsfbox{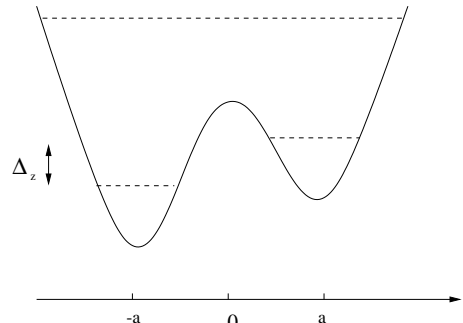}}
\caption{
Schematic representation of a DWP with two \\minima. 
The dotted lines represent the lower 
energy levels \\and $\Delta_z$ is the level anisotropy.
}
\label{fig:DWP}
\end{figure} 
}

\section{The Model}
The partition function describing an heavy particle 
moving in a DWP  and  interacting with an electron bath can be written 
in the following way:
\be 
{\cal Z}=\int D[q(\tau)] D[\bar\psi({\bf x},\tau)]D[\psi({\bf x},\tau)]
\re^{- S_E[q,\bar\psi,\psi]}.
\label{z1}
\ee
where $S_E$ is the Euclidean action
\be
S_E=\int \rd 
\tau {\cal L}
\ee
with
\be
{\cal L}=  {\cal L}_{at}+{\cal L}_{el}+\delta {\cal L}.
\label{l}
\ee
Through this Section we will consider only $T=0$, an extension to finite 
temperature is trivial and will be discussed at the end of 
Sec.~\ref{sec:mapping}.

In Eq.~(\ref{l}) 
 ${\cal L}_{at}$ describes the dynamics of the heavy particle and is given
by
\be
{\cal L}_{at}=\frac{M}{2}\left ( \frac{dq}{d\tau} \right )^2+V(q),
\label{latom}
\ee
where $q(\tau)$ represent the position of the atom 
and $V(q)$ is the confining potential, which we assume to have two minima 
like in Fig.~\ref{fig:DWP}, and assume that the level anisotropy is absent,
$\Delta_z=0$, or more precisely, 
\be
V(q)=V(-q).
\label{symmetry}
\ee 
Note that here and in the following 
the potential enters the Lagrangian with a minus sign with respect to the usual
definition of the Lagrangian since we work in 
Euclidean space.

The bare electronic Lagrangian has the form
\be
{\cal L}_{el}=\int \rd^3 {\bf x} \bar \psi({\bf x},\tau)
\partial_\tau \psi({\bf x},\tau)+H_0
\ee
with 
\be
H_0=\int \rd^3 k  \; 
\epsilon_{\bf k} \bar{\psi}({\bf k})\psi({\bf k}) 
\ee
where
$\psi({\bf x})$ is the electron field
\be
\psi({\bf x})
=\int \frac{\rd^3  k}{(2\pi)^3} \re ^{i {\bf k} \cdot {\bf x} }\psi({\bf k}).
\ee 
For the moment we do not include spin, as
it is only a  spectator. It will be reintroduced at the end 
where it will play the role of the channel index.
We chose a general
density-density interaction between the electrons and the atom of 
the form
\be
\delta{\cal L}=
\delta H=\lambda \int \rd^3 x \bar \psi({\bf x},\tau) \psi({\bf x},\tau)  
\delta ({\bf x}-q(\tau))
\label{dH}
\ee
which, as explained in the introduction, we believe contains all the essential 
physics.

Given the symmetry of the problem it is convenient to expand the electron 
field in partial waves
\be
\psi({\bf r})=\sum_{l,m}\psi_{l,m}=\sum_{l,m}\int_0^\infty
 \frac{dk}{2\pi} R_{k,l}(r)
Y_{l,m} \left (\frac{\bf r}{r} \right ) \psi_{k,l,m}
\ee
where $Y_{l,m}$ are spherical harmonics and $R_{k,l}=2k j_l(kr)$, with
$j_l(x)=\sqrt{\pi/2x} J_{l+1/2}(x)$ being Bessel functions. 
>From the condition 
\be
k_F a \ll 1,
\label{cond.kfa}
\ee
where $a$ is the distance between two minima of the potential, it 
follows that only the first two harmonics, $l=0$, and $l=1$, $m=0$,
strongly interact with the heavy 
particle. Then we can approximate the electron field as
\bea
\psi({\bf r})
&\simeq& Y_{0,0} \left (\frac{\bf r}{r} \right ) 
\int_{0}^{\infty} 
\frac{dp}{2\pi} 
\psi_{+}(p) R_{p,0}(r)
\nonumber\\     
&+&
Y_{1,0} \left (\frac{\bf r}{r} \right )
\int_{0}^{\infty} 
\frac{dp}{2\pi} 
\psi_{-}(p) R_{p,1}(r),
\label{psi}
\eea
where we have introduced $\psi_+(k)=\psi_{k,0,0}$ and $\psi_-(k)=\psi_{k,1,0}$.
The low energy properties are determined by
electron states close to the
Fermi surface, $p \ll k_F$,
so that we can introduce fields smooth on the scale $1/k_F$ 
\be
\psi_{\pm}(x)=\int_{p \ll k_F}\frac{dp}{2\pi} e^{ipx} \psi_{\pm}(p).
\label{1dfermions}
\ee
These fields
determine the asymptotic
behavior of the three dimensional field at $rk_F \gg 1$,
thus specifying the scattering matrix for the states close to the
Fermi energy shell. In Eq.~(\ref{psi}) we suppressed all the higher
angular harmonics not scattered by the heavy particle.
Then, using the asymptotic behaviour  of the Bessel functions for 
$rk_F \gg 1$ and Eq.~(\ref{1dfermions}), we can rewrite Eq.~(\ref{psi}) as
\bea
\psi({\bf r})
&\simeq& \frac{1}{r} Y_{0,0}\left (\frac{\bf r}{r} \right )
\left[\psi_{+}(r)e^{ik_Fr} - \psi_{+}(-r)e^{-ik_Fr}
\right]\nonumber\\
&+&
\frac{1}{r} Y_{1,0} \left (\frac{\bf r}{r} \right ) 
\left[\psi_{-}(r)e^{ik_Fr} + 
\psi_{-}(-r)e^{-ik_Fr}
\right]\nonumber \\
&=&
\frac{1}{i\sqrt{4\pi}r}
\left[\psi_{+}(r)e^{ik_Fr} - \psi_{+}(-r)e^{-ik_Fr}
\right]\nonumber\\
&+& \frac{\sqrt 3 z}{\sqrt{4\pi} r^2}
\left[\psi_{-}(r)e^{ik_Fr} + 
\psi_{-}(-r)e^{-ik_Fr}
\right],
\label{asymp}
\eea
where we have chosen the 
direction of the motion of the heavy particle restricted along the 
$z$ axis. 
This leaves us with an effective one dimensional problem with two species
of electrons. These are left movers on the entire axis as 
in Eq.~(\ref{asymp}) or, equivalently, 
one can introduce left and right movers on the half line.
We have  chosen the former approach for technical convenience.
%
%
With the same accuracy we can linearize the spectrum in the vicinity
of the Fermi surface: $\epsilon_{k}=v_F(k -k_F)$ and
$v_F|k-k_F|\ll  D$,  where $D$ is the energy scale
smaller than the Fermi energy.

Before we proceed, let us note that 
there have been some suggestions \cite {ZZ2} 
that the electron-hole 
asymmetry of the original electronic band 
can lead to an enhancement of the 
Kondo temperature. 
These effects will not be taken into account in this Section,
and the corresponding discussion is relegated to the Appendix. 
As we will explicitly show  there,
they  can be  introduced perturbatively and do
{\sl not} change qualitatively our results. 

For the linearized spectrum, the bare part of the 
fermionic Hamiltonian takes the
standard form
\be
H_0= iv_F\int dx 
\left[\bar{\psi}_+\partial_x\psi_+ +
\bar{\psi}_-\partial_x\psi_-\right]. 
\label{h0}
\ee
Substitution of Eq.~(\ref{psi}) into Eq.~(\ref{dH}) 
yields
\bea
\delta &H&=\frac{\lambda}{4\pi} 
\int \frac{dp_1}{2\pi}\frac{dp_2}{2\pi}
\label{dh1}
\label{dH1}
\\
&\times&\left[
\bar{\psi}_{+}(p_1) R_{k_F+p_1,0}[q(\tau)]
+\sqrt{3}\bar{\psi}_{-}(p_1) R_{k_F+p_1,1}[q(\tau)]
\right]
\nonumber\\
&\times&
\left[
{\psi}_{+}(p_2) R_{k_F+p_2,0}[q(\tau)]
+\sqrt{3}{\psi}_{-}(p_2) R_{k_F+p_2,1}[q(\tau)]
\right].
\nonumber
\eea


Now we neglect $p_{1,2}$ in comparison with the Fermi momentum and use
the condition
(\ref{cond.kfa}) to expand Eq.(\ref{dh1})
in $k_F q(\tau)$ up to the second order. The smallness of this parameter is
guaranteed by Eq.~(\ref{cond.kfa})  together with the 
fact that the dynamics of the heavy particle is dominated
by $q(\tau) \simeq \pm a$.  We obtain
\begin{mathletters}
\label{dh2}
\bea
\delta H&=&\delta H_0 + \delta H_q 
\nonumber\\
\delta H_0&=&\frac{\lambda k_F^2}{\pi}
\left[\left(1- \frac{k_F^2a^2}{3}\right)
\bar{\psi}_+{\psi}_+
+\frac{k_F^2a^2}{3}\bar{\psi}_-{\psi}_-
\right] \label{dh2a} \\
\delta H_q &=& \frac{\lambda k_F^2}{\pi}
\left[\frac{k_Fq(\tau)}{\sqrt{3}}
\left(\bar{\psi}_+{\psi}_- + \bar{\psi}_-{\psi}_+\right)
\right.\nonumber\\
&+& 
\left.
\frac{k_F^2\left[a^2 - q^2(\tau)\right]}{3}
\left(
\bar{\psi}_+{\psi}_+ -  \bar{\psi}_-{\psi}_-
\right)\right],
\label{dh2b}
\eea
\end{mathletters}
from which it is clear that mixing of higher harmonics would have
been of order ${\cal O}((k_F a)^3)$. 
Here and in the following  we  use the short hand notation
\[
{\psi} \equiv \frac{{\psi}(+0) + {\psi}(-0)}{2},
\]
whenever there is no spatial integration. 
The Hamiltonian (\ref{dh2a}) describes the electron scattering on a static
potential created by the 
heavy particle smeared between potential minima.
This term does not have any dynamics and may be eliminated by a unitary
transformation. On the other hand, the
term (\ref{dh2b}) describes the excitations
of the electron system by the moving heavy particle, and should be treated
carefully.

To get rid of the Hamiltonian (\ref{dh2a}), we perform the transformation
%
%
\begin{eqnarray}
&&\psi_\pm(x) \to \psi_\pm(x)e^{i{\rm sgn}(x) \delta_{\pm}},
\quad
\bar{\psi}_\pm(x) \to \bar{\psi}_\pm(x)e^{-i{\rm sgn}(x) \delta_{\pm}}
\nonumber
\\
&&\tan\delta_+ =  \frac{\lambda k_F^2}{\pi v_F} 
\left(1- \frac{k_F^2a^2}{3} \right), \quad
\tan\delta_- =  \frac{\lambda k_F^4a^2}{3 \pi v_F},
\label{transform}
\end{eqnarray}
in Eqs.~(\ref{h0}) and (\ref{dh2}). After this transformation
the term $H_0+\delta H_0$ acquires the form (\ref{h0}), and
Eq.~(\ref{dh2b}) becomes
\begin{eqnarray}
\frac{\delta H_q}{v_F}&=&
2\pi \Lambda_z \frac{q(\tau)}{4a} 
\left[\bar{\psi}_+{\psi}_-+\bar{\psi}_-{\psi}_+\right]
\label{dh6}\\
&+&2\pi \Lambda_x\frac{a^2-q^2(\tau)}{2 a^2}
\left[\bar{\psi}_+{\psi}_+-\bar{\psi}_-{\psi}_-\right]
\nonumber\\
&+& 2 \pi \Lambda_\rho\frac{a^2-q^2(\tau)}{2 a^2}
\left[\bar{\psi}_+{\psi}_++\bar{\psi}_-{\psi}_-\right],
\nonumber
\end{eqnarray}
where the dimensionless constants $\Lambda_{x,z,\rho}$ are  
\begin{eqnarray}
&&\Lambda_z=  \frac{ 2 k_Fa}{\pi \sqrt{3}}  
\sin (\delta_++\delta_-),\nonumber\\
&&\Lambda_x= \frac{1}{2\pi}\left (\frac{k_Fa}{\sqrt{3}}\right )^2 
\sin (\delta_++\delta_-)
\frac{\cos^2 \delta_+ +\cos ^2\delta_-}{\cos \delta_+ \cos \delta_-},
\nonumber\\
&&\Lambda_\rho = \frac{1}{2\pi}\left (\frac{k_Fa}{\sqrt{3}} \right )^2 
\sin (\delta_++\delta_-)
\frac{\cos^2 \delta_+ -\cos ^2\delta_-}{\cos \delta_+ \cos \delta_-}.
\label{7}
\end{eqnarray}
 It follows from the condition (\ref{cond.kfa}) that 
\be
\Lambda_x,\;
\Lambda_\rho \ll \Lambda_z \ll 1,
\label{relation}
\ee
independently of the value of the coupling constant $\lambda$.
Similar conclusions about the values of the coupling constants
 was reached in Ref.~\cite{KP2}.

Introducing
the pseudo-spin notation
\begin{equation}
\mbox{\boldmath $\psi$}= \pmatrix{\psi_+ \cr \psi_-},\quad
\bar{\mbox{\boldmath $\psi$}}= \pmatrix{\bar{\psi}_+ ,\ \bar{\psi}_-},
\end{equation}
and choosing the Pauli matrices in pseudo-spin space in the following 
representation
\begin{equation}
\tau_z =\pmatrix{ 0 &1\cr 1&0}, \quad  
\tau_y =\pmatrix{ 0 &i\cr -i&0}, \quad \tau_x =\pmatrix{ 1 & 0\cr 0&-1},
\end{equation}
the expressions  
(\ref{h0}) and (\ref{dh6}) can be rewritten in the more compact form
\begin{eqnarray}
\frac{{H_0+\delta H_q}}{v_F} &=& 
i\int dx 
\mbox{\boldmath $\bar{\psi}$}\partial_x\mbox{\boldmath ${\psi}$}
+ 2\pi\Lambda_z Z(\tau)
\mbox{\boldmath $\bar{\psi}$}\frac{\tau_z}{2}\mbox{\boldmath ${\psi}$} 
\label{8}\\
&+&
2\pi \Lambda_x X(\tau)
\mbox{\boldmath $\bar{\psi}$}\frac{\tau_x}{2}\mbox{\boldmath ${\psi}$}
+
\pi \Lambda_\rho X(\tau)
\mbox{\boldmath $\bar{\psi}$}\mbox{\boldmath ${\psi}$},
\nonumber
\end{eqnarray}
where we introduced the short hand notation
\be
Z(\tau) = \frac{q(\tau)}{2a}; \quad
X(\tau) = \frac{a^2-q^2(\tau)}{ a^2}.
\label{coordinates}
\ee
One can see from the Hamiltonian (\ref{8}) that $v_F$ can be removed by the
proper rescaling of the time coordinates. In what follows, we will
set $ v_F = 1$.

To calculate the partition function it is convenient
to introduce the $U(1)$ (charge)  and  $SU(2)$ (pseudo-spin) currents
\begin{equation}
J=\mbox{\boldmath $\bar{\psi}$}\mbox{\boldmath ${\psi}$},
\quad \vec{J}=
\mbox{\boldmath $\bar{\psi}$}\frac{\vec{\tau}}{2}\mbox{\boldmath ${\psi}$}
\label{currents}
\end{equation}
that satisfy the commutation relations
\begin{mathletters}
\label{comm}
\begin{eqnarray}
&&[J(x),\ J(0)]=i\frac{\delta^\prime(x)}{\pi};
\label{comm1}
\\
&&[J_j(x),\ J_k(0)]=i
\frac{\delta_{jk}\delta^\prime(x)}{4\pi}
+ i \epsilon^{jkl}\delta(x)
J_l(x);
\label{comm2}
\\
&& [J(x),\ J_k(y)]=0.
\nonumber
\end{eqnarray}
\end{mathletters}
The pseudo-spin current commutation relations (\ref{comm2}) 
define  the  $SU(2)$ 
Kac-Moody algebra at level 1 ($SU(2)_1$).
In terms of the currents (\ref{currents}) the Hamiltonian (\ref{8}) 
takes the Sugawara form
\begin{mathletters}
\label{sugawara}
\begin{eqnarray}
{H_0}&=& \int dx \left \{
\frac{\pi}{2} J^2 + \frac{2 \pi}{3} \vec{J}^2 \right \}
\label{12}
\\
\delta H &=& 2\pi \Lambda_z Z (\tau) J_z 
+ 2\pi \Lambda_x X(\tau) J_x 
+ \pi \Lambda_\rho X(\tau) J.
\label{122old}
\end{eqnarray}
Pseudo-spin and charge degrees of freedom
are separated and in the bare part, $H_0$,
one recognizes  the free charge boson model and
the  $SU(2)_1$ Wess-Zumino-Novikov-Witten (WZNW) model. For a future use,
we generalize Eq.~(\ref{122old}) by introducing one more coupling $\Lambda_y$:
\bea
\frac{\delta H}{2\pi } &=& 
\Lambda_z Z (\tau) J_z 
+  \Lambda_x X(\tau) J_x + \Lambda_y Y(\tau) J_y
+ \frac{\Lambda_\rho X(\tau) J}{2}, \nonumber\\
Y(\tau)&=&-i\frac{\tau_{tun}}{2a}\frac{dq}{d\tau},
\label{122}
\eea
where $i$ is introduced in front of the first time derivative because
we work with the imaginary time.
The parameter $\tau_{tun}$ has the dimensionality of
time and is introduced to make $\Lambda_y$ dimensionless. It is defined
rigorously in the next section, however, its exact meaning is not
important here.
In the original model at high energies $\Lambda_y=0$, however, in our analysis,
this 
coupling will be generated in the higher order perturbation theory,
as shown in the  next section. 
\end{mathletters}

As first step in the process of mapping our problem to the LGM
we need to integrate out the fermionic degrees of freedom. In the framework
of the formalism  that we have chosen here, this requires 
to construct the Action corresponding to the Hamiltonian (\ref{sugawara}). 
A general procedure to do it is provided by non-Abelian bosonization
and leads to the WZNW
 Action (see for instance \cite{book.GNT}).
For simplicity we chose a different approach that uses Abelian bosonization. 
Despite the fact that the $SU(2)$ is a non-Abelian group,
the use of Abelian bosonization in this case is made possible  
by the fact that the $SU(2)_1$ WZNW model has central charge equal to one. 
Introducing two bosonic fields for the charge and pseudo-spin 
degrees of freedom in the standard way 
\begin{mathletters}
\label{bosonization}
\bea
J&=&\frac{1}{\sqrt{\pi}}\partial_x \phi_c;
\label{bosonization1}\\
J_z&=&\frac{1}{\sqrt{4\pi}} \partial_x \phi_s ; \nonumber \\
J_+ &\equiv& J_{x} + i J_{y}
= \frac{D }{2\pi } \re^{-i \sqrt{8\pi} \phi_s} \nonumber\\
J_ -&\equiv& J_{x} -   i J_{y}
= \frac{D}{2\pi} \re^{i \sqrt{8\pi} \phi_s}
\label{bosonization2}
\eea
\end{mathletters}
we can write the electronic Lagrangian density in the bosonic form
\be
{\cal L}_b={\cal L}_b^0+\delta H
\ee
where 
\bea
{\cal L}_b^0&=&\frac{1}{2}\sum_{\rho=c,s} \left [ 
i\partial_\tau \phi_\rho
\partial_x \phi_\rho + \left ( \partial_x \phi_\rho \right )^2 
\right ] ,
\eea
and $\delta H$ can be easily obtained  substituting Eqs.~(\ref{bosonization})
into (\ref{122}). In the bosonization formulas (\ref{bosonization2}), 
$D$ is an arbitrary cut-off required to make the correlation functions 
in the bosonic theory finite. It can be fixed by requiring
that the correlation function of vertex operators in this 
theory is given by
\be
\label{cut-off}
\langle \exp(i\sqrt{4\pi}\phi_s(x)) \exp(i\sqrt{4\pi}\phi_s(0))\rangle=
\frac{1}{D x}.
\ee

After bosonizing we can integrate out 
the electronic degrees of freedom 
in (\ref{z1}) and rewrite it as
\be
{\cal Z}=\int D[q(\tau)] \; \re^{-\int \rd \tau {\cal L}_{at}}
{\cal Z}_\rho[q(\tau)] {\cal Z}_{xzy}[q(\tau)]
\label{zeff}
\ee
where ${\cal L}_{at}$ is defined in Eq.~(\ref{latom}) and the 
product ${\cal Z}_\rho[q(\tau)] {\cal Z}_{xzy}[q(\tau)]$
is nothing but the electronic determinant for the given path of heavy
particle $q(\tau)$. It has the following form
\end{multicols}
\widetext

\begin{mathletters}
\label{expansion}
\be
{\cal Z}_\rho=
\Bigg\langle \exp\left(- \pi \int d\tau
\Lambda_\rho X(\tau) J \right)
\Bigg\rangle
=\exp\left( \frac{\Lambda_\rho^2}{2}
\int d\tau_1d\tau_2
\frac{X(\tau_1)X(\tau_2)}{(\tau_1-\tau_2)^2}
\right); 
\label{expansion.zrho}
\ee
\bea
{\cal Z}_{xzy}&=&
\sum_m \left(\frac{1}{m!}\right)^2 \left 
(2\pi\right )^{2m} 
 \int 
\left[\prod_{j=1}^m\rd \tau_j^+d \tau_j^-
R_+(\tau_j^+)R_-(\tau_j^-)
\right]
\Bigg\langle  \prod_{j=1}^m J_+(\tau^+_j)J_-(\tau^-_j)
\exp\left(- 4\pi \int d\tau\Lambda_z Z(\tau) J_z\right)\Bigg
\rangle
\nonumber\\
& =&
\sum_m \left(\frac{1}{m!}\right)^2 \left 
(2\pi  \right )^{2m}
\nonumber \\ &\times &\int 
\left[\prod_{j=1}^m\rd \tau_j^+d \tau_j^-
R_+(\tau_j^+)R_-(\tau_j^-)
\right]
\exp\left[{\Lambda_z} 
\int d\tau {Z(\tau)}\sum_{j=1}^{m} \left(\frac{1}{\tau-\tau_j^+}
-\frac{1}{\tau-\tau_j^-}\right)\right]
{\cal Z}_z \Bigg\langle \prod_{j=1}^m J_+(\tau^+_j)J_-(\tau^-_j)
\Bigg\rangle \nonumber 
\\
&=&
\sum_m  \left(\frac{1}{m!}\right)^2 
\int 
\left[\prod_{j=1}^m\rd \tau_j^+d \tau_j^-
R_+(\tau_j^+)R_-(\tau_j^-)
\right]
\nonumber \\
&\times&\exp\left[{\Lambda_z} 
\int d\tau Z(\tau)\sum_{j=1}^{m} \left(\frac{1}{\tau-\tau_j^+}
-\frac{1}{\tau-\tau_j^-}\right)\right] {\cal Z}_z
\prod_{i,j}^m \frac{1}{(\tau_i^+-\tau_j^-)^2}
\prod_{i>j}^m (\tau_i^+-\tau_j^+)^2 (\tau_i^--\tau_j^-)^2,
\label{expansion.xz}
\eea
where 
\[
R_{\pm}(\tau) \equiv 
\frac{\Lambda_x}{2}X(\tau) \pm \frac{\Lambda_y}{2}Y(\tau),
\]
functions $Z(\tau)$, $X(\tau )$, and $Y(\tau )$ are defined in
Eqs.~(\ref{coordinates}), (\ref{122}),
and
\be
{\cal Z}_z=\Bigg\langle  \exp\left(- 2\pi \int d\tau
\Lambda_z {Z(\tau)} J_z\right)\Bigg\rangle
= \exp\left( \frac{\Lambda_z^2}{2}
\int d\tau_1d\tau_2
\frac{Z(\tau_1)Z(\tau_2)}{(\tau_1-\tau_2)^2}
\right).
\label{expansion.z} 
\ee
\end{mathletters} 

\begin{multicols}{2}
In Eqs.~(\ref{expansion}) averaging over the bosonic fields is defined
as
\[
\langle\dots\rangle =\int D[\phi(x,\tau)] \re^{\int \rd \tau {\cal L}_b^0} ...
\]

%
In the next Section we will use a semi-classical approximation to map
(\ref{zeff}) into the partition function of a one-dimensional gas
of logarithmically interacting particles.
 This will enable us to establish  a non-perturbative relationship
between our original problem and the Kondo model.

\section{Mapping to a logarithmic gas model}
\label{sec:mapping}

The main goal of this section is to map the model of the previous section
to the logarithmic gas model. A similar approach for the TLS was done in 
\cite{vzz,zznvz}. As usual, for such kind of mapping,
the first step is to identify the logarithmically interacting objects.
We will show in Sec. \ref{sec:3a} that,
for the present problem,  they are 
instanton-antiinstanton 
configurations of the tunneling problem\cite{instantons}.
In order to find the upper cut-off for the logarithmic gas problem
${\cal D}$
which plays the role of the prefactor in the Kondo temperature, we
calculate in Sec.~\ref{sec:3b} the interaction between the instantons.
As we will see, it turns out that there is an intrinsic short distance cut-off,
that makes the theory regular, and is
given by the size of the instanton. It is 
analogous to the lattice constant for 2D melting problem, or
to the vortex core size for superfluid films. Then a 
rough estimate for ${\cal D}$ is given by this quantity.
To define the cut-off with the higher accuracy we find the coupling
constants corresponding to  the chosen value of ${\cal D}$ by microscopic
calculation of the energy of the logarithmic gas model. 
The relation between the scale ${\cal D}$ and the coupling constants
will be shown to have the form of 
\bea
h({\cal D})= h\ \left[1 + {\Lambda_z^2} \left(\ln {\cal D}\tau_{tun}
+{\cal O}(1)\right)\right]; \nonumber\\
\Lambda_y({\cal D})= -\Lambda_x\Lambda_z  
\left(\ln {\cal D}\tau_{tun} +{\cal O}(1)\right).
\label{firstlog}
\eea
On the other hand, Eq.(\ref{firstlog}), can be interpreted as the
result of the first iteration of the renormalization group equations
for the Kondo problem with $\Lambda_y \ll \Lambda_x,\Lambda_z$:
 \be
\frac{d h({\cal D})}{d \ln {\cal D} }=  h\ {\Lambda_z^2}; \quad 
\frac{d \Lambda_y({\cal D})}{d \ln {\cal D} }= -\Lambda_x\Lambda_z  
\label{RG}
\ee
It means that 
the cut-off ${\cal D}$ could be chosen arbitrary, provided
that the coupling constants are adjusted accordingly, as it is
well-known for any logarithmical problem. Accordingly, we will chose
the cut-off ${\cal D}$ to have the Hamiltonian of the most simple
form:
\be
 \Lambda_y({\cal D}) = 0.
\label{DEQ}
\ee 
This equation exclude any ambiguity in the definition of ${\cal D}$ 
and will allow
us to determine the cut-off even with numerical coefficient for
a wide class of the potentials $V(q)$.

\subsection{Dilute instanton gas approximation}
\label{sec:3a}

The partition function (\ref{zeff}) 
can be calculated using a semi-classical approximation,
which amounts to considering only small fluctuations around the
stationary 
points of the Action, which correspond to the 
Euler-Lagrange (EL) equations for the 
problem. 
A rough condition for the applicability of the semi-classical approximation
is $S_{inst}\gg 1$, where $S_{inst}$ is the classical action corresponding
to the trajectory connecting two extrema of the potential $V(q)$.

To first approximation 
we can calculate the stationary points using the EL equations 
for the bare problem 
\be
{M}\frac{\rd ^2 q}{\rd \tau^2}=V'[q],
\label{el0}
\ee
which corresponds to the  equation of 
motion for a classical 
particle in the potential  minus $V[q]$. This
approximation can be improved by taking into account 
the modification of the optimal trajectory due to the electrons,
however, it introduces only parametrically
small changes of the coupling constants 
(because the shape of the classical solution is a rigid mode)
and does not change  any  of our conclusions.
Beside the trivial solution $q(\tau)=\pm a$, Eq.~(\ref{el0}) 
also admits solutions  of the form
\be
q(\tau)=\pm a f(\tau-\tau_i)
\label{kink}
\ee
(kink  and anti-kink respectively)
with $f(\pm \infty)=\pm 1$.  
This solutions are called instantons because they 
produce an almost instantaneous blimp in the Lagrangian.
The action is invariant under translations 
of the instanton center $\tau_i$, which reflects the time translation 
invariance of the original Lagrangian (\ref{latom}). 
The instanton is characterized by its bare action
\be
S_{inst}=\int \rd \tau {\cal L}_{at}[af(\tau)],
\label{instaction}
\ee
and by the tunneling time
\be
\tau_{tun}=\int d\tau \big[1-f^2(\tau)\big].
\label{ttun}
\ee
By construction, the integrand is non-zero only within the core of the
instanton, so $\tau_{tun}$ has the meaning of the size of the core.

In what follows, we will write the explicit results for the model potential
\be
V(q)=g(a^2-q^2)^2.
\label{pot}
\ee
even though our considerations by no means are restricted for such
potential.
For the potential (\ref{pot}), one easily finds
\be
S_{inst}=\frac{M^2\omega^3}{12g};\quad
f(\tau)=\tanh\frac{\omega\tau}{2}; \quad 
\tau_{tun} =\frac{4}{\omega},
\label{actionmodel}
\ee
where $\omega^2={8 g a^2}/{M}$ is the frequency of the harmonic
oscillation in the extrema of the potential.

{\begin{figure}[ht]
\epsfxsize=6cm
\centerline{\epsfbox{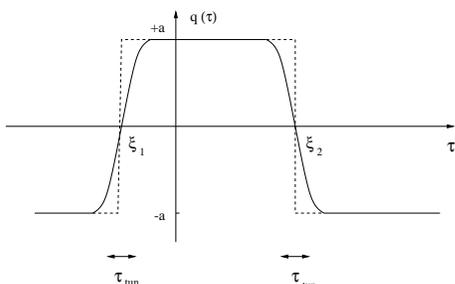}}
\caption{
Schematic form of the instanton solution (\ref{kink}) \\ (solid line).
The dotted line represents
instantaneous spin  \\ flip in the Kondo problem.
}
\label{fig:instantons}
\end{figure} 
}

The partition function must sum over 
multi-instanton solutions satisfying initial and final conditions. 
Since the instanton core is small,  it is a good 
approximation to consider multi-instanton solutions 
(see Fig.~\ref{fig:instantons}) as if instanton and 
anti-instantons were dilute (dilute instanton gas 
approximation), which requires
$|\tau_i-\tau_j| \gg \tau_{tun}$ (this assumption is completely justified 
for 
$S_{inst} \gg 1$ \cite{instantons}). 
Then, in the absence of the  
interaction of heavy particle with electrons,
a general 
trajectory 
starting and ending at point $-a$
can be written as
\begin{equation}
q_n(\tau) = - a \Big[1+ \sum_{j=1}^{2n}
(-1)^jf(\tau-\xi_j) \Big],
\label{14}
\end{equation} 
where the
positions of the kinks $\xi_{2j+1}$ 
anti-kinks  $\xi_{2j}$ 
are subject to constraints
\begin{equation}
\xi_j  < \xi_{j+1}. 
\label{15}
\end{equation} 

The action corresponding to a configuration 
with $n$ instanton-antiinstanton pairs 
with the exponential accuracy is 
\be
\int \rd \tau {\cal L}_{at}\Big[q_n(\tau)\Big] =
2n S_{inst},
\label{dilgasaction}
\ee
i.e.  kinks would not interact with each other if the
heavy particle were isolated from the electron system. 

In the dilute instanton gas approximation, the low-temperature partition
function of an isolated heavy particle can be rewritten also in a form
\be
{\cal Z_0}= \sum_{n=0}^\infty h^{2n}\prod_{j=1}^{2n} \int_0^{1/T} d\xi_j=
\cosh \frac{h}{T}
\label{Zfree}
\ee
Hereinafter, integrals over $\xi_j$ are calculated with the 
constraint (\ref{15}).
The tunneling splitting between two lowest levels of the heavy particle is 
estimated as
\be
h = \kappa \frac{1}{\tau_{tun}}\sqrt{S_{inst}}e^{-S_{inst}} 
\ll \frac{1}{\tau_{tun}}
\label{tunnsplitt}
\ee
The numerical factor $\kappa$ comes from the instanton determinant
due to the integration around
the saddle point\cite{instantons}. For potential (\ref{pot}),
one finds $\kappa=\sqrt{24/\pi} \approx 2.76$.

Closing the subsection, we discuss the limits of the applicability of the 
dilute instanton approximation. If the potential $V(q)$ does not have 
singular points the main condition of the applicability is $S_{inst} \gg 1$.
We emphasize, that this condition has nothing to do with the number of
levels localized in each well. In particular, numerical solution of
the Schr\"{o}dinger equation with the potential (\ref{pot}), shows that the
third level in DWP lies larger than the maximum of the potential already
at $S_{inst} \leq 6.06$, (at this point $h\tau_{tun}=10^{-2}$), whereas
the dilute instanton approximation breaks down at  $S_{inst} \simeq 2.8$,
(when the second level crosses the maximum of the potential).

\subsection{Interaction between instantons}
\label{sec:3b}

The purpose of this subsection is to establish the form of interaction
between instantons and anti-instantons due to the dynamics of the electron
system. To do so we substitute Eq.(\ref{14}) into Eq.~(\ref{zeff}).
The functional integration is reduced to the integration around small 
oscillations about the optimal trajectory (\ref{14}). The calculation of such 
instanton determinant is standard 
\cite{instantons}and may be performed without taking 
into account the electrons, provided that the condition (\ref{relation}) holds.
As the result we find 
\be
{\cal Z}= \sum_{n=0}^\infty h^{2n}\prod_{j=1}^{2n} \int
d\xi_j
{\cal Z}_\rho\big[\{\xi\}_{j=1}^{2n}\big] 
{\cal Z}_{xzy}\big[\{\xi\}_{j=1}^{2n}\big], 
\label{whereto}
\ee
where ${\cal Z}_{\rho,xzy}\big[\{\xi\}_{j=1}^{2n}\big]$ are the integrals
(\ref{expansion}) calculated over $q(\tau)$ given by Eq.~(\ref{14}). 
These
functions  
depend on the positions
of the kinks and therefore produce interactions between them.
In what follows, we analyze this interaction in details. 

Substituting Eq.~(\ref{14}) into Eq.~(\ref{coordinates}) and neglecting
the 
exponentially small overlap of the instanton cores one can rewrite $Z(\tau)$
and $X(\tau)$ as 
\bea
Z(\tau) &=& \frac{1}{2}\Big[-1+ \sum_{j=1}^{2n}
(-1)^jf(\tau-\xi_j) \Big]\nonumber\\
X(\tau) &\approx&
\sum_{j=1}^{2n}\eta_1 (\tau-\xi_j),
\quad \eta_1 \equiv 1-f^2.
\label{coor1}
\eea
It is clear that the function $X(\tau)$ has peaks at the core of instantons
and vanishes exponentially otherwise.

Our strategy now is to substitute Eq.~(\ref{coor1}) into
Eqs.~(\ref{expansion}) and perform simplifications using the fact that
the instanton gas is dilute. This can be easily done for the terms
(\ref{expansion.zrho}) and (\ref{expansion.z}) while 
Eq.~(\ref{expansion.xz}) requires more work. 
For Eq.~(\ref{expansion.zrho}) we find\cite{footnote}
\bea
&&\int d\tau_1d\tau_2
\frac{X(\tau_1)X(\tau_2)}{(\tau_1-\tau_2)^2}
= 2n \delta S_0 
+\sum_{i \neq j}^{2n}
U_{0}(\xi_i-\xi_j); 
\nonumber\\
&&\delta S_0 \equiv U_0(0);
\quad  
U_0(\tau)= {\rm Re}
\int 
\frac{d\tau_1d \tau_2 
\ \eta_1(\tau_1)\eta_1(\tau_2)}{(\tau_1-\tau_2-\tau + i0)^2}.
\nonumber\\
\label{zrho1}
\eea
One can see from Eqs.~(\ref{zrho1}) and (\ref{coor1}) 
that $U_0(\tau)$ decays 
rapidly: $U_0(\tau) \simeq (\tau_{tun}/\tau)^2$ at  $\tau > \tau_{tun}$.
This short range interaction between the instantons is not important
for the dilute instanton gas and we can neglect it. Thus,
for a configuration of $n$ kink - anti-kink pairs, 
Eq.~(\ref{expansion.zrho}) takes the form
\be
{\cal Z}_\rho = \exp(2 n\Lambda_\rho^2\delta S_0)
\label{zrho2},
\ee
which is just an independent
renormalization of the action for each kink. Such renormalization is nothing
but the polaronic effect for the tunneling.

Next, we substitute Eq.~(\ref{coor1}) into Eq.~(\ref{expansion.z}) and,
omitting an irrelevant constant term, 
we obtain 
\bea
&&\int d\tau_1d\tau_2
\frac{Z(\tau_1)Z(\tau_2)}{(\tau_1-\tau_2)^2}
= 2n \delta S_1
+\sum_{i \neq j}^{2n}(-1)^{i+j}
U_{1}(\xi_i-\xi_j);
\nonumber\\
&&
U_1(\tau)=
\int d\tau_1d\tau_2 
{\eta_2(\tau_1)\eta_2(\tau_1)}
\ln {\cal D}\Big|{\tau_1-\tau_2-\tau}\Big|;
\nonumber\\
&&\eta_2(\tau) \equiv \frac{d f}{2 d\tau},\quad 
\delta S_1 \equiv U_1(0),
\label{zz1}
\eea
where energy scale ${\cal D}$ is introduced here to make the 
argument of the logarithmic interaction dimensionless: it does not
enter into the expression for the total action.
In the following we will see that the natural choice 
is ${\cal D}=1/\tau_{tun}$,
where the tunneling time is defined in Eq.~(\ref{ttun}), however, we will
keep  ${\cal D}$ as an
independent energy scale for pedagogical reasons.

One can see, {\em e.g.} from Fig.~\ref{fig:instantons}, that function
$\eta_2(\tau)$ decays rapidly at $\tau > \tau_{tun}$.  Therefore,
the interaction between kinks, $U_1(\tau)$, is logarithmic at $\tau \gg
\tau_{tun}$ and it saturates at $\tau \simeq \tau_{tun}$.  This is
nothing but the manifestation of the usual orthogonality
catastrophe\cite{Orthogonality}, where the high-energy cut-off is
determined by the dynamics of the heavy particle.

Substituting (\ref{zz1}) into Eq.~(\ref{expansion.z}) we find that this
part of the partition function is equivalent to that of the classically
interacting gas with $n$ positive and $n$ negative particles:
\bea
&&{\cal Z}_z = \exp(n\Lambda_z^2\delta S_1)
\exp(-{\cal H}_1)\label{zz0}\\
&& {\cal H}_1 = -
\frac{\Lambda_z^2}{2} \sum_{i \neq j}^{2n}(-1)^{i+j}
U_{1}(\xi_i-\xi_j).
\nonumber
\eea

To begin the 
manipulations with  the contribution (\ref{expansion.xz}), we first
study the simplest, $m=1$, term to illustrate the principle, and then
switch to the higher order terms. 
We rewrite the prefactor in Eq.~(\ref{expansion.xz}) as
\end{multicols}
\widetext
\bea
&&\frac{ R_+(\tau_+)R_-(\tau_-)}{\left(\tau_+ - \tau_- \right)^2}
 =
\sum_{j=1}^{2n}
\frac{\eta_+^j(\tau_+-\xi_j)\eta_-^j(\tau_- -\xi_j)}
{\left(\tau_+ - \tau_-\right)^2}
+ \sum_{i\neq j}^{2n}
\frac{\eta_+^j(\tau_+-\xi_j)\eta_-^i(\tau_- -\xi_j)}
{\left(\tau_+ - \tau_-\right)^2}
\approx
\label{xz1}
\\
&&
\sum_{j=1}^{2n}
\frac{\eta_+^j(\tau_+-\xi_j)\eta_-^j(\tau_- -\xi_j)}
{\left(\tau_+ - \tau_-\right)^2}
+ \sum_{i\neq j}^{2n}
\frac{\eta_+^j(\tau_+-\xi_j)\eta_-^i(\tau_--\xi_i)}
{\left(\xi_i - \xi_j\right)^2}, \quad
\eta_{\pm}^j(\tau) = \frac{\Lambda_x}{2} \eta_1(\tau)
\mp (-1)^j \frac{\Lambda_y}{2} \tau_{tun}\eta_2(\tau) 
\nonumber
\eea 
where in the last line we used once again  the fact that the
instanton gas is dilute, and functions $\eta_{\pm}(\tau)$ decay
exponentially outside the core of the instantons. 
The first term in the last line
of Eq.~(\ref{xz1}) keeps times $\tau_{\pm}$ close to each other,
and, as we will see, will produce the renormalization of the action
for a single kink. The second term will give rise to the interaction
between kinks.
\begin{multicols}{2}

For the exponent in Eq.~(\ref{expansion.xz}) one obtains
\bea
&&\int d\tau Z(\tau) \left(\frac{1}{\tau-\tau^+}
-\frac{1}{\tau-\tau^-}\right)
\nonumber\\
&&\quad
= \sum_{j=1}^{2n}(-1)^j 
\left[
U_2(\xi_j - \tau_+) - U_2(\xi_j - \tau_-)
\right],
\nonumber\\
&&U_2(\tau)=
\int d\tau_1 
\eta_2(\tau_1)
\ln {\cal D}\Big|{\tau_1-\tau}\Big|,
\label{xz2}
\eea 
where we introduced the arbitrary cut-off $D$ for the same purpose as
in Eq.~(\ref{zz1}). Similarly to $U_1(\tau)$, the potential
$U_2(\tau)$ is logarithmic at $\tau \gg \tau_{tun}$ and it saturates
at $\tau \simeq \tau_{tun}$.

>From  Eqs.~(\ref{xz1}) and (\ref{xz2}), we can easily rewrite, within the
dilute instanton approximation, the 
$m=1$ contribution
to ${\cal Z}_{xzy}$ as\cite{footnote}
\bea
&&\frac{{\cal Z}_{xzy}^{(m=1)}}{ {\cal Z}_z   } =   
2n \Bigg[ I_x \frac{\Lambda_x^2}{4}+ I_y \frac{\Lambda_y^2}{4} 
+{\cal O}(\Lambda_x\Lambda_y\Lambda_z)
\Bigg]
+ K_1
; \label{xz3}\\
&&I_x={\rm Re}
\int d\tau_1d\tau_2 \frac{\eta_1(\tau_1)\eta_1(\tau_2)}
{\left(\tau_1-\tau_2 + i0\right)^2},
\nonumber\\
&&I_y=\tau_{tun}^2{\rm Re}
\int d\tau_1d\tau_2 \frac{\eta_2(\tau_1)\eta_2(\tau_2)}
{\left(\tau_1-\tau_2 + i0\right)^2},
\nonumber\\
&&4 K_1=\tau_{tun}^2
\sum_{i\neq j}^{2n}
\frac{\left[\tilde{\Lambda}_x - (-1)^i \tilde{\Lambda}_y
\right]
\left[
\tilde{\Lambda}_x + (-1)^j \tilde{\Lambda}_y\right]
}
{\left(\xi_{ij}\right)^2},
\nonumber\\
&&\times \exp
\left\{\Lambda_z
{\sum\limits_{k =1}^{2n}}^\prime
(-1)^k \left[U_2(\xi_{ik})-
U_2(\xi_{jk}) \right]
\right\},
\nonumber
\eea
where we used the short hand notation
\be
\xi_{ij} \equiv \xi_{i}-\xi_{j}
\ee
for the distance between the instantons, and 
${\sum}^\prime$ means that terms involving $\xi_{ii}$
are excluded.
 Deriving $K_1$ from
 Eq.~(\ref{xz3}),
we used the properties $\int d\tau \eta_1(\tau)=\tau_{tun}$,
and $\int d\tau \eta_2(\tau)=1$.
The dimensionless coupling constants $\tilde{\Lambda}_{x,y}$
entering into Eq.~(\ref{xz3}) are defined as
\bea
\tilde{\Lambda}_x=
\Lambda_x - \Lambda_y  \Lambda_z
\int {\rd\tau} \eta_2(\tau) U_2(\tau) + {\cal O}(\Lambda_z^2\Lambda_x)
\nonumber\\
\tilde{\Lambda}_y=
\Lambda_y - \Lambda_x \Lambda_z
\int \frac{\rd\tau}{\tau_{tun}} \eta_1(\tau)
U_2(\tau) + {\cal O}(\Lambda_z^2\Lambda_y)
\label{L}
\eea

The factor $K_1$ can be also rewritten in terms of the partition function
of the classical logarithmic gas. In order to do so, we
relate to any $j$th instanton the additional charge
\be
\mu_j= -1,\ 0,\ 1.
\label{q}
\ee
As we will see immediately, the 
physical meaning of the ``neutral'' kink, $\mu_j=0$ is the tunneling of
the heavy particle without excitations of the electron system,
where as ``charged'' kinks  $\mu_j=\pm 1$ describe the electron assisted
tunneling.
Using the  notation introduced in (\ref{q}) and 
 the fact that asymptotic
behavior of the potential $U_2(\tau)$ is logarithmic, we rewrite
Eq.~(\ref{xz3}) as: $K_1=K_{m=1}$, where the more general quantity 
$K_{m}$ is given by
\bea
&&K_m\Big(\{\xi_j\}_{j=1}^{2n}\Big)
=
\sum_{\{\mu_j\}_{j=1}^{2n}}\Gamma\big(\{\mu_j\}\big)
\exp\left(-{\cal H}_2\right) \nonumber \\
&&{\cal H}_2 = - \sum_{i\neq j}^{2n}
U_2(\xi_{ij})
\left[\mu_j\mu_i + (-1)^i \mu_j \Lambda_z\right]
\nonumber\\
&&
\Gamma\big(\{\mu_j\}\big)=
\prod_{j=1}^{2n}\left(1-\mu_j (-1)^j \frac{\tilde\Lambda_y}
{\tilde\Lambda_x}\right)
\left(\frac{\tilde\Lambda_x {\cal D} \tau_{tun}}{2}\right)^{|\mu_j|}
. \label{H2}
\eea

The  summation over all the configurations of $\{\mu_j\}$
is subject to the condition of charge neutrality 
\be
\label{cn}
\sum_{j=1}^{2n}\mu_j = 0
\ee
and to the $m$th order perturbation theory constraint
\be
\sum_{j=1}^{2n}|\mu_j| = 2m.
\label{cm}
\ee 

Equations (\ref{xz3}) and (\ref{cm}) allow for very natural generalization
to the higher order terms. 
For a configuration containing $n$ kinks and $n$ anti-kinks
we wish to keep only terms of the order of $(n\Lambda_x^2)^m$
and neglect the terms which scale like  $(n\Lambda_x^4)^m$ etc.
It amounts to the neglecting in a product
$\prod_{j=1}^{m}X(\tau_j^+)X(\tau_j^-)$ configurations that contain
more than two kinks coinciding. 
We employ this approximation to extend Eq.~(\ref{xz3}) for general $m$.
To get rid of the combinatorial factor $1/(m!)^2$, we
impose additional constraints on the integration $\tau_j^+ <
\tau_{j+1}^+$, and $\tau_j^- <
\tau_{j+1}^-$.
The result then acquires the form\cite{disclaimer}
\be
\frac{{\cal Z}_{xzy}^{(m)}}{ {\cal Z}_z   } =   
\frac{\Lambda_x^{2m}}{4^m}
\sum_{k=0}^m \frac{1}{k!} 
(2n I)^k  K_{m-k}\Big(\{\xi_j\}_{j=1}^{2n}\Big),
\label{xz5}
\ee
where the combinatorial factor takes care of the ordering
of the paired 
kinks  and  the factor $K_{m}$
is given by Eq.~(\ref{H2}) with the
constraints (\ref{cn}) and (\ref{cm}).

The total contribution for $2n$ instantons is obtained by 
summation of
all the orders of perturbation theory:
\bea
&&\frac{{\cal Z}_{xzy}}{ {\cal Z}_z   } =
\frac{1}{ {\cal Z}_z   }
\sum_{m=0}^\infty {\cal Z}_{xzy}^{(m)} \label{xz6}
\nonumber\\
 &&
=\sum_{m=0}^\infty
\sum_{k=0}^m
\frac{(\Lambda_x^{2}n I)^{k}
\left(\Lambda_x^{2}/4\right)^{m-k} K_{m-k}}{k!}
\nonumber\\
 &&
=
e^{n\Lambda_x^{2}I}
\sum_{m=0}^\infty
{\left(\Lambda_x^{2}/4\right)^{m}K_{m}}. 
\label{xzfinal}
\eea

We now substitute
Eqs.~(\ref{xzfinal}), (\ref{zz0}), and (\ref{zrho2}) into
Eq.~(\ref{whereto}).
As  result  one finds
\bea
{\cal Z}= 
\sum_{n=0}^\infty 
\tilde{h}^{2n}
\sum_{\{\mu_j\}}
\Gamma\big(\{\mu_j\}\big)\int {d\xi_j}
e^{-{\cal H}},
\label{zfinal}\\
\nonumber
\eea
where the integration is performed with the constraint (\ref{15}),
$\Gamma\big(\{\mu_j\}\big)$ are defined in Eq.~(\ref{H2}),
and the summation over charge configurations $\{\mu_j\}$ is performed
with the neutrality condition (\ref{cm}).
Level splitting due to the spontaneous tunneling  renormalized by the
interaction with electrons (compare it with (\ref{tunnsplitt})) is
\bea
\label{SSS}
\tilde{h}=h \exp\left(
\frac{s\Lambda_\rho^2U_0(0)}{2} 
+\frac{s\Lambda_z^2U_1(0)}{2} +
\frac{s\Lambda_x^{2}I_x}{2}
\right) \\
\approx h \left(1+
\frac{s\Lambda_\rho^2U_0(0)}{2} 
+\frac{s\Lambda_z^2U_1(0)}{2} +
\frac{s\Lambda_x^{2}I_x}{2}
\right).
\nonumber
\eea
with  entries defined in Eqs.~(\ref{7}), (\ref{zrho1}), (\ref{zz1})
and (\ref{xz3}), and $s=1$, and we put $\Lambda_y=0$.

The energy of the classical logarithmic gas 
${\cal H}={\cal H}_1+{\cal H}_2$ is found as
\be
{\cal H} =
 - \sum_{i\neq j}^{2n}
U_2(\xi_{ij})
\left[\mu_j\mu_i + (-1)^i \mu_j 
\Lambda_z + \frac{(-1)^{i+j}\Lambda_z^2}{2}\right], 
\label{heff}
\ee
where we neglected the difference between the potentials $U_2$
and $U_1$ from Eqs.~(\ref{xz2}) and (\ref{zz1})
for  $\xi_{ij}$ much larger than the size of the
core of the instantons.

So far we have considered only spinless electrons.
The real spin is trivially included.
We notice that there the electron spin commutes with the Hamiltonian,
and therefore the fermionic determinant for spin $1/2$ electrons
is factorized onto product of two fermionic determinants for each spin.
It results in the replacement
\[
{\cal Z}_{\rho, xzy} \to \left[{\cal Z}_{\rho, xzy}\right]^2
\] 
in Eqs.~(\ref{zeff}) and~(\ref{whereto}). Such replacement is taken
into account by
introducing an additional 
``spin'' for the kink, $\sigma^j = \uparrow,\downarrow$. By simple
repeating
of all of the consideration of this Subsection, one finds
instead of Eq.~(\ref{zfinal}) 
\be
{\cal Z}= 
\sum_{n=0}^\infty 
\tilde{h}^{2n}
\sum_{\{\mu_j\},\ \{\sigma_j\}}
\Gamma\big(\{\mu_j\}\big)\int {d\xi_j}
e^{-{\cal H}},
\label{zfinalspin}
\ee
with the Hamiltonian 
\bea
&&{\cal H} =
 - \sum_{i\neq j}^{2n}
U_2(\xi_{ij})
\Bigg\{\delta_{\sigma_i\sigma_j}\left[\mu_j\mu_i + 
(-1)^i \mu_j \Lambda_z\right]
\nonumber\\
&&\hspace*{2cm}
 + \frac{(-1)^{i+j}\Lambda_z^2}{2}\Bigg\},
\label{hspin}
\eea
and $\Gamma\big(\{\mu_j\}\big)$ are defined in Eq.~(\ref{H2}).
The expression for the action (\ref{SSS}) should be used
with $s=2$. This constitute the  result of the
mapping of our original problem to a 
one-dimensional classical gas of charged particles interacting via the 
potential $U_2(r)$ given by Eq.~(\ref{xz2})\cite{finiteT}. 
The fugacity and the interaction potentials have been calculated starting 
from a completely microscopic theory.

Energies of the logarithmic gas explicitly depend on the high energy
cutoff ${\cal D}$, see Eq.~(\ref{xz2}), and so do the coupling
constants (\ref{L}) and (\ref{SSS}). It is easy to see that the
dependence of the coupling constants of cut-off indeed has the form
of Eq.~(\ref{firstlog}) (more formal argument that $\Lambda_y$ is indeed 
analogous to the corresponding coupling constant for the Kondo problem
will be given in the next section after Eq.~(\ref{1000a})). 
Indeed, with the help of
Eq.~(\ref{xz2}), we rewrite Eqs.~(\ref{L}) and (\ref{SSS}) as
\begin{mathletters}
\bea
&&\tilde{\Lambda}_y= - \Lambda_x \Lambda_z
\left[\ln {\cal D}\tau_{tun}
+ \alpha_y \right];\\
&&\tilde{h}= h \left\{1+
\Lambda_z^2
\left[\ln {\cal D}\tau_{tun}+
\alpha_h \right]
+
\Lambda_\rho^2U_0(0) 
 +
{\Lambda_x^{2}I_x}
\right\};\\
&&\alpha_y=\int \frac{\rd\tau_1\rd\tau_2}{\tau_{tun}} 
\eta_1(\tau_1)\eta_2(\tau_2)
\ln \Big|\frac{{\tau_1-\tau}}{\tau_{tun}}\Big|; \label{gammay}\\
&&\alpha_h=\int {\rd\tau_2\rd\tau_2} 
\eta_2(\tau_1)\eta_2(\tau_2)
\ln \Big|\frac{{\tau_1-\tau}}{\tau_{tun}}\Big|; \label{gammah}
\eea
\end{mathletters}
Because $\int d\tau \eta_1(\tau)=\tau_{tun}$, and
$\int d\tau \eta_2(\tau)=1$, the
parameters $\alpha_{h,y}$, $I_x$ and $U_0$ 
can take only numerical values of the order of unity not depending of
the cut-off ${\cal D}$. We now define ${\cal D}$ in accordance with
the rule (\ref{DEQ}). We obtain
\be
{\cal D} = \frac{1}{\tau_{tun}}e^{-\alpha_y},
\label{result}
\ee
which together with Eqs.~(\ref{gammay}), (\ref{kink}), (\ref{ttun}),
(\ref{coor1}), (\ref{zz1}) solves the problem of the relation of
the high-energy cut-off with the form of the instanton solution of
the heavy-particle dynamics in the double well potential.
Parameter $h$, in its turn acquires the form
\be
\tilde{h}= h \left[1+
\Lambda_z^2
\left(
\alpha_h -\alpha_y\right)
+
\Lambda_\rho^2U_0(0) 
 +
{\Lambda_x^{2}I_x}
\right].
\ee
For illustration purposes we calculate the numerical value of the
constant for the potential (\ref{pot}). Using explicit form of the
instanton solution (\ref{actionmodel}), one immediately finds:

\begin{eqnarray}
&&U_0(0)=I_x= -\frac{12}{\pi^2}\zeta(3)\approx -1.461,\nonumber\\
&&\alpha_y=\alpha_h = 
\frac{1}{4} \ln \left(\frac{2 e^{{\bf C}+1}}{\pi}\right)
\approx 0.281. \label{coeffmodel}
\end{eqnarray}
where $\zeta(3)$ is the Riemann Zeta function, and ${\bf C} \approx 0.577$
is the Euler constant.

Closing the section, we write down the result for the logarithmic
gas model with the cut-off result. For $\Lambda_y=0$, Eq.~(\ref{H2}) 
for factors $\Gamma\big(\{\mu_j\}\big)$ are simplified and one obtains
from Eqs.~(\ref{zfinalspin}), (\ref{hspin})
\be
{\cal Z}= 
\sum_{n=0}^\infty 
\tilde{h}^{2n}\sum_{\{\mu_j\},\ \{\sigma_j\}}
\prod_{j=1}^{2n}
\left(\frac{\Lambda_x e^{-\alpha_y}}{2}\right)^{|\mu_j|}
\int {d\xi_j}
e^{-{\cal H}},
\label{zfinalspin1}
\ee
with the Hamiltonian 
\bea
&&{\cal H} =
 - \sum_{i\neq j}^{2n}
\ln ({\cal D}\xi_{ij})
\Bigg\{\delta_{\sigma_i\sigma_j}\left[\mu_j\mu_i + 
(-1)^i \mu_j \Lambda_z\right]
\nonumber\\
&&\hspace*{2cm}
 + \frac{(-1)^{i+j}\Lambda_z^2}{2}\Bigg\},
\label{hspin1}
\eea

The results above will enable us in the next Section
to make connection between the 
microscopic parameters of the model of tunneling centers and the 2CK model.

\section{Relation to the two channel Kondo model and impossibility of
the strong coupling limit}

A similar procedure can be done for the 2CK model with real spin \cite{AY,FGN}.
In our notation Eq.~(\ref{kondo}) acquires the form of Eq.~(\ref{sugawara}) 
with 
 (\ref{122}) of the form
\be
\delta H = 4\pi \lambda_z S_z J_z 
+ 4\pi \lambda_x S_x J_x + 4\pi \lambda_y S_y J_y + 2 h S_x,
\label{1Kondo}
\ee
where we included constant $\lambda_y$ for the sake of generality.

We extended the results of Refs.~\onlinecite{AY,FGN} 
to include the presence of a 
transverse magnetic field by the repeating 
all the steps of the previous section.
The only difference is that the spin flips are instantaneous 
(see Fig.~\ref{fig:instantons}), so
that the high-energy cut-off is determined by the electronic scale $D$.
 We obtained the partition function 
for the 2CK model (\ref{1Kondo}) in the form (\ref{zfinalspin})
as
\be
{\cal Z}= 
\sum_{n=0}^\infty 
h^{2n}\sum_{\{\mu_j\},\{\sigma_j\}}
\int d\xi_j
e^{-{\cal H}_{Kondo}},
\label{1000}
\ee
with
\be
\Gamma\big(\{\mu_j\}\big)=
\prod_{j=1}^{2n}\left(1-\mu_j (-1)^j \frac{\lambda_y}
{\lambda_x}\right)
\left ( \frac{\lambda_x D}{2 h} \right )^{|\mu_j|}
\label{1000a}
\ee
and 
\bea
{\cal H}_{Kondo}&=& - \sum_{i\neq j}^{2n}
\Bigg\{\delta_{\sigma_i\sigma_j}\left[\mu_j\mu_i + 
(-1)^i \mu_j 2\lambda_z\right]
\nonumber\\
&&\hspace*{2cm}
 + (-1)^{i+j}2 \lambda_z^2\Bigg\} \ln|D \xi_{ij}/\hbar|
\label{2000}
\eea
Again  the integration is performed with the constraint (\ref{15})
and the summation over charge configurations $\{\mu_j\}$ is subject to 
the neutrality condition (\ref{cn}). 
The cut-off $D$ is the same cut-off as in Eq.~(\ref{cut-off}), 
or more precisely
\[
\lim_{\tau \to \infty}
\Big[\ln D\tau
+2\pi^2 {\rm Re} \int_0^\tau d\tau_1\int_0^\tau  d\tau_2
\langle J_x(\tau_1) J_x(\tau_2)\rangle
\Big] =0.
\]

Let us now compare the logarithmic gas models  for tunneling centers
(\ref{zfinalspin}), (\ref{zfinalspin1}) 
and for the Kondo problem (\ref{1000}). First of all, direct comparison 
of Eq.~(\ref{1000a}) with Eq.~(\ref{H2}) shows that constant
$\Lambda_y$  is exactly equivalent to the coupling constants of Kondo
model indeed and Eq.~(\ref{RG}) follow.
Next, we put $\lambda_y=0$ in Eq.~(\ref{1000a}) and compare the
result with Eq.~(\ref{zfinalspin1}). 
We immediately find
that two models become equivalent upon the following 
identification of the parameters
\bea 
&&D \leftrightarrow   {\cal D} = \frac{e^{-\alpha_y}}{ \tau_{tun}}
\nonumber\\
&&{h} \leftrightarrow \tilde{h}\approx h \nonumber \\
&&\lambda _x   \leftrightarrow    \Lambda_x 
\left(\frac{h}{e^{\alpha_y}\cal D}\right) = \Lambda_x 
\left({h}\tau_{tun}\right)
\nonumber \\
&&\lambda_z  \leftrightarrow   \frac{\Lambda_z}{2},
\label{rel}
\eea
where the tunneling time is defined in Eq.~(\ref{ttun}), and numerical
constant $\alpha_y$ is defined in Eq.~(\ref{gammay}) for the arbitrary
DWP and calculated for model (\ref{pot}) in Eq.~(\ref{coeffmodel}).

We note that the equivalence  between the tunneling impurity model and the 
2CK model for times larger then $\tau_{tun}$,  is non-perturbative in 
the sense that it is established at any order in perturbation theory.
We reiterate, that the mapping has to be performed with 
 account of all of the excited states of the
movable atom.

>From the  relationships (\ref{rel}) 
it is easy to show that the Kondo temperature 
is always smaller then $h$ ($\Delta_z=0$ in our model) and then the 2CK 
regime can never be reached. Indeed, inserting Eqs.~(\ref{rel})
into (\ref{tk}), one has 
\be
\frac{T_K}{ h}  = \re^{-\alpha_y}\Lambda_x
 \left ( \frac{\Lambda_x }{ \Lambda_z} 
\right )^\gamma \left( h\tau_{tun}\right)^\gamma,
\label{tk2}
\quad
\gamma=\frac{1}{\Lambda _z}-\frac{1}{2}.
\ee
>From the condition (\ref{cond.kfa}), and (\ref{7}) it follows that
\be
\Lambda_x \ll 1, \quad
\gamma\gg 1, \quad \frac{\Lambda_x}{\Lambda_z} \ll 1.
\ee
>From Eq.~(\ref{tunnsplitt}), we have usual relation for the tunneling splitting
of two levels:
\[
h{\tau_{tun}} \ll 1.
\]

Together with Eq.~(\ref{tk2}), this implies that 
\be
T_K/h \ll 1
\ee
and then {\em the strong coupling regime cannot
be reached}.

\section{discussion}

In this paper we considered a general model describing a tunneling impurity 
moving in a double well potential and embedded in a metal. 
The main motivation for this work  was to 
provide a conclusive answer to the  
question weather it is possible to observe the strong coupling regime
of the two-channel Kondo 
model in such a system. In order to answer this question
one has to find the correct relationship between
the microscopic parameters of the tunneling impurity problem and the 
coupling constants of the effective two-channel
Kondo  model. Previous results showed that
the two-level system
is not a good starting point because all the excited states of 
the impurity play an essential role.
In order to take into account all the excited states of the 
double well potential problem  we used a 
different approach. We mapped the tunneling impurity model into a 
one-dimensional logarithmic gas model using 
a semi-classical (dilute instanton) approximation to describe 
the dynamics of the impurity. 
Since the same mapping can be done  for the two-channel Kondo
model, we obtain a general relationship
between the coupling constants of the two 
models. This relationship is obtained taking into account all the 
excited levels of the heavy particle and is valid to any order in 
perturbation theory.
We demonstrated that, in the effective two-channel Kondo model
(\ref{kondo}),   
the values of the coupling
constant $\lambda_x$  and the transverse magnetic field $h$,
are intrinsically related, and can be never
considered independently. This fact, together with the existence of an
intrinsic high-energy cut-off in the theory, $1/\tau_{tun}$, conspire in 
such a way that  the Kondo temperature is always smaller then 
the effective magnetic field $h$. Then
the strong 
coupling fixed point of the two-channel Kondo
model can {\em never} be reached in this 
system. 
This results are valid for any form of the double well potential
and are robust against 
specific properties of the  electron system.

\acknowledgments
We would like to thank B. Altshuler for  important discussions at the early 
stage of this work. We are also  grateful to K. Held and J. von Delft for 
discussions and to L. Glazman, N. Prokof'ev and A. Tsvelik 
for reading the manuscript and
interesting comments.
One of us (I.A.) was supported by Packard foundation. Work in Lancaster 
University was partially funded by EPSRC-GR/R01767.

\end{multicols}

\widetext
\appendix
\section{Effect of the electron-hole asymmetry}
\label{ap:A}

In this appendix we present the general scheme
for the calculation of the parameters of the effective model 
(\ref{zfinalspin})
for a general electron spectrum without involving the 
linearized approximation (\ref{h0}) from the very beginning. 
Our motivation for doing so is to provide framework in which
further discussion (quite futile to our opinion) of the role of
electron-hole asymmetry should be performed.

Integration over the fermionic fields in Eq.~(\ref{z1}) gives the
formal result
\be
\int D[\bar\psi({\bf x},\tau)]D[\psi({\bf x},\tau)]
\re^{- S_E[q,\bar\psi,\psi]}
= 
\re^{- S_{el}[q(\tau)] -\int d\tau {\cal L}_{at}(q,\dot{q}) },
\label{a1}
\ee
where the effect of the electrons on the atom is described by
\be
- S_{el}[q(\tau)]= \label{a2}
{\rm Tr}
\ln \Bigg[-
\frac{\p}{\p\tau} - \hat\xi({\bf p}) - V\Big(i\frac{\p}{\p {\bf p}}
+\hat{\mbox{\boldmath $\Omega$}}({\bf p})
-{\bf q}(\tau)\Big)
\Bigg],
\ee
where ${\bf p}$ is the quasi-momentum of the electrons,  $\xi({\bf p})=
{\rm diag} [\xi_j({\bf p})]$ is 
the spectrum of Bloch electrons, and
$V({\bf r})$
is the potential of the interaction of the atom with the electrons,
which can be of more general form than the local interaction
(\ref{dH}). Finally, $\hat{\mbox{\boldmath $\Omega$}}({\bf p})$
is the standard\cite{Textbook} non-diagonal component of the coordinate operator
in the basis of the Bloch functions and it describes
the inter-band scattering due to the atomic potential.
In Eq.~(\ref{a2}) and thereafter all the energies are counted from
the Fermi level.

It will be convenient for us to express all the quantities in terms of
the solution of the scattering problem on immobile atom. In order to
do so, we perform the unitary transformation in Eq.~(\ref{a2}) as
\[
{\rm Tr}
\ln \Big[\dots
\Big]= {\rm Tr}
\ln \Big[ \re^{i{\bf q}(\tau){\bf p}}
\dots
\re^{-i{\bf q}(\tau){\bf p}}
\Big]
\]
and obtain from Eq.~(\ref{a2})
\bea
- S_{el}[q(\tau)]= \label{a3}
{\rm Tr}
\ln \Big[-
{\p}_{\tau} + i{\bf p}
\dot{\bf q}(\tau) - \hat\xi({\bf p}) - \hat{V} 
\Big],
\quad \hat{V}\equiv  V\Bigg(i\frac{\p}{\p {\bf p}}
+\hat{\mbox{\boldmath $\Omega$}}({\bf p})\Bigg).
\eea

Our goal now is to expand Eq.~(\ref{a3}) in powers of $\dot{q}$,
and relate the expansion coefficients to the coupling constants of the
low energy theory (\ref{zfinalspin}). To facilitate such an expansion
we introduce the Matsubara Green functions and $\hat{T}$ matrix of
immobile atom
\bea
&&\hat{G}_0(i\varep_n, {\bf p}) = \frac{1}{i\varep_n -\hat\xi({\bf p})};
\nonumber\\
&&\hat{G}(i\varep_n, {\bf p}_1,
 {\bf p}_2) = \left[\frac{1}{i\varep_n -\hat\xi({\bf p}) -
\hat{V}}\right]_{{\bf p}_1
 {\bf p}_2} 
= \hat{G}_0(i\varep_n, {\bf p}_1)
\Big[\delta_{{\bf p}_1
 {\bf p}_2} + \hat{T}(i\varep_n, {\bf p}_1,
 {\bf p}_2) \hat{G}_0(i\varep_n, {\bf p}_2)
\Big];
\nonumber\\
&&
\hat{T}(i\varep_n, {\bf p}_1,{\bf p}_2) = 
\Bigg[ \hat{V}\left(1- \hat{G}_0(i\varep_n) \hat{V}\right)^{-1}
\Bigg]_{{\bf p}_1{\bf p}_2} 
\label{a4} 
\eea
with $\varep_n=\pi T(2n+1)$ being the fermionic Matsubara frequency,
and their retarded and advanced counterparts
\be
\hat{G}^{R(A) }_0(\varep) = \hat{G}_0(\varep \pm i 0);
\quad \hat{T}^{R(A) } (\varep) =  \hat{T}(\varep \pm i 0).
\label{a5}
\ee

We note the identities
\bea
&&\hat{G}_0(i\varep_n, {\bf p}) -
\hat{G}_0(i\varep_n+ i\omega_n, {\bf p}) = 
i \omega_n \hat{G}_0(i\varep_n, {\bf p})\hat{G}_0(i\varep_n+ i\omega_n, {\bf p});
\nonumber\\
&& 
\hat{T}(i\varep_n, {\bf p}_1, {\bf p}_2) -
\hat{T}(i\varep_m, {\bf p}_1, {\bf p}_2)=
\int \frac{d^3 p_3}{(2\pi)^3} 
\hat{T}(i\varep_n, {\bf p}_1, {\bf p}_3)\nonumber\\
&&
\ \ \times
\Big[
\hat{G}(i\varep_n,{\bf p}_3)
-\hat{G}(i\varep_m,{\bf p}_3)
\Big]
\hat{T}(i\varep_m, {\bf p}_3, {\bf p}_2)
\label{a6}
\eea
where $\omega_m=2\pi Tm$ is the bosonic Matsubara frequency. 
Hereafter the momentum integration is performed within the first
Brillouin zone.

Using Eq.~(\ref{a4}) we rewrite Eq.~(\ref{a3}) in the form of
linked cluster expansion
\be 
S_{el}[q(\tau)]= 
\sum_{m=1}^\infty
\frac {S_{el}^{(m)}}{m};
\quad
S_{el}^{(m)}=
 {
{\rm Tr}
\left[ - i\dot{\bf q}{\bf p}
\hat{G}
\right]^m},
\label{a7}
\ee
where we omitted the term independent of $q(\tau)$ and
all of the multiplications should be understood in the matrix sense.

Before we proceed  we notice that in the absence of the impurity potential
quasi-momentum ${\bf p}$ is an integral of motion. Therefore, the coupling
to  ${\bf p}$ of the force 
with non-zero Matsubara frequency has no effect
independently of the spectrum $\xi({\bf p})$. Using the fact  
that
\be
\int_0^{1/T} d\tau \dot {\bf q} =0,
\label{a8}
\ee
we find natural result that
\be
{\rm Tr}
\left[\dot{\bf q}{\bf p}
\hat{G}_0
\right]^m = 0.
\label{a9}
\ee

Now we perform the actual calculation of the expansion 
(\ref{a6}). According to Eq.~(\ref{a7}), the first order term
vanishes, $S_{el}^{(1)}=0$. Using Eqs.~(\ref{a5}) and (\ref{a9}) one finds 
for the second order term
\bea
&&S_{el}^{(2)} = 
T\sum_{\omega_n}\sum_{\alpha\beta=x,y,z}\omega_n^2
{q}_\alpha(\omega_n){q}_\beta(-\omega_n) \Pi^{\alpha\beta}(\omega_n)
\label{a10}\\
&&\Pi^{\alpha\beta}=
-T\sum_{\varep_n}
\int \frac{d^3 p}{(2\pi)^3}p^\alpha p^\beta
{\rm Tr}\Bigg\{
\hat{G}_0(i\varep_n, {\bf p})\hat{T}(i\varep_n, {\bf p}, {\bf p})
\hat{G}_0(i\varep_n, {\bf p})\hat{G}_0(i\varep_n+i\omega_n, {\bf p}) 
 + \omega_n \to -\omega_n
\Bigg\}
\nonumber\\
&&
-T\sum_{\varep_n}
\int \frac{d^3 p}{(2\pi)^3}
\frac{d^3 k}{(2\pi)^3}
p^\alpha k^\beta
{\rm Tr}\Bigg\{
\hat{G}_0(i\varep_n+i\omega_n, {\bf p})
 \hat{T}(i\varep_n+i\omega_n, {\bf p}, {\bf k})
\hat{G}_0(i\varep_n+i\omega_n, {\bf k})\hat{G}_0(i\varep_n, {\bf k})
 \hat{T}(i\varep_n, {\bf k}, {\bf p})\hat{G}_0(i\varep_n, {\bf p})
\Bigg\}.
\nonumber
\eea
with
\be
 {\bf q}(\omega_n) = \int_0^{1/T} d\tau  {\bf q}(\tau)e^{i\omega_n\tau}.
\label{a11}
\ee

With the help of Eq.~(\ref{a6}), one rewrites Eq.~(\ref{a10}) 
\bea
&&\Pi^{\alpha\beta}=
- \frac{T}{\omega_n^2}\sum_{\varep_n}
\int \frac{d^3 p}{(2\pi)^3}
\frac{d^3 k}{(2\pi)^3}
\Big(p^\alpha p^\beta - p^\alpha k^\beta\Big)
\label{a12}
{\rm Tr}\Big\{\hat{G}^*({\bf p})
\hat{T}(i\varep_n+i\omega_n, {\bf p}, {\bf k})
\hat{G}^*({\bf k})\hat{T}(i\varep_n, {\bf k}, {\bf p}) 
\Big\};
\nonumber\\
&&\hat{G}^*({\bf p}) \equiv \hat{G}_0(i\varep_n+i\omega_n, {\bf p}) - 
\hat{G}_0(i\varep_n, {\bf p})
\nonumber
\eea
Performing the standard trick with the replacement of summation over
$\varep_n$ to the integration, we obtain
\bea
&&\Pi^{\alpha\beta}=
- \frac{1}{\omega_n^2}
\int\frac{d\varep}{\pi}
\tanh \frac{\varep}{2T}
\int \frac{d^3 p}{(2\pi)^3}
\frac{d^3 k}{(2\pi)^3}
\Big(p^\alpha p^\beta - p^\alpha k^\beta\Big)
\nonumber\\
&&
\ \
\times
{\rm Im}
\Bigg\{
{\rm Tr}\Big[\hat{G}^+({\bf p})
\hat{T}(\varep+i|\omega_n|, {\bf p}, {\bf k})
\hat{G}^+({\bf k})\hat{T}^R(\varep, {\bf k}, {\bf p}) 
\Big]
-{\rm Tr}
\Big[\hat{G}^-({\bf p})
\hat{T}(\varep+i|\omega_n|, {\bf p}, {\bf k})
\hat{G}^-({\bf k})\hat{T}^A(\varep, {\bf k}, {\bf p}) 
\Big]
\Bigg\}
;
\eea
where
\bea
&&\hat{G}^+({\bf p}) \equiv \hat{G}_0(\varep+i|\omega_n|, {\bf p}) - 
\hat{G}_0^R(\varep, {\bf p})
\nonumber\\
&&\hat{G}^-({\bf p}) \equiv \hat{G}_0(\varep+i|\omega_n|, {\bf p}) - 
\hat{G}_0^A(\varep, {\bf p})
\label{a13}
\eea

At that point it is important to emphasize that
$|{\bf q}(\omega_n)|$ decays exponentially at large frequencies
for the saddle point solution (\ref{kink}),
$|{\bf q}(\omega_n)|\propto \re^{-|\omega_n|\tau_{tun}}$. The fluctuations
around the saddle point are also suppressed at large frequencies due
to the  kinetic energy in Eq.~(\ref{latom}). The Green functions
in Eq.~(\ref{a13}) are analytic functions of energy except the branch
cut at the real axis. It allows one to expand over $|\omega_n|$ in Eq.~(\ref{a13}):
\bea
&&\Pi^{\alpha\beta}=-2
\int\frac{d\varep}{\pi}
\tanh \frac{\varep}{2T}
\int \frac{d^3 p}{(2\pi)^3}
\frac{d^3 k}{(2\pi)^3}
\Big(p^\alpha p^\beta - p^\alpha k^\beta\Big)\times
\nonumber\\
&&
\Bigg\{
\frac{1}{| \protect\omega_n |}
{
\frac{\p}{\p\varep}
{\rm Tr}\
\Big[{\rm Im}\hat{G}_0^R(\varep, {\bf p})\Big]
\hat{T}^R(\varep, {\bf p}, {\bf k})
\Big[{\rm Im}\hat{G}_0^R(\varep, {\bf k})\Big]
\hat{T}^A(\varep, {\bf k}, {\bf p})}
- \frac{1}{2}{\rm Im}{\rm Tr}\
\frac{\p \hat{G}_0^R(\varep, {\bf p})}{\p \varep}
\hat{T}^R(\varep, {\bf p}, {\bf k})
\frac{\p\hat{G}_0^R(\varep, {\bf k})}{\p \varep}
\hat{T}^R(\varep, {\bf k}, {\bf p})
\nonumber\\
&&
\ \ 
+ \frac{\p}{\p\varep}
{\rm Re}{\rm Tr}\
\Big[{\rm Im}\hat{G}_0^R(\varep, {\bf k})\Big]
\hat{T}^A(\varep, {\bf k}, {\bf p})
\Bigg[
\Big[{\rm Im}\hat{G}_0^R(\varep, {\bf p})\Big]
\p_\varep\hat{T}^R(\varep, {\bf p}, {\bf k})
+ 
\Big[{\rm Re}\p_\varep \hat{G}_0^R(\varep, {\bf p})\Big]
\hat{T}^R(\varep, {\bf p}, {\bf k})
\Bigg]
\Bigg\}
\nonumber\\
\label{a14}
\eea
Substituting Eq.~(\ref{a14}) into Eq.~(\ref{a10}) we find
\begin{mathletters}
\label{a15}
\be
S_{el}^{(2)} = T\sum_{\omega_n}\sum_{\alpha\beta=x,y,z}
{q}_\alpha(\omega_n){q}_\beta^*(\omega_n)
\Bigg(|\omega_n| R^{\alpha\beta} + \omega_n^2 Q^{\alpha\beta}
\Bigg).
\ee
Here
\bea
&&R^{\alpha\beta}=2\pi
\int{d\varep}
\frac{d}{d\varep}\tanh \frac{\varep}{2T}
\int \frac{d^3 p}{(2\pi)^3}
\frac{d^3 k}{(2\pi)^3}
\Big(p^\alpha p^\beta - p^\alpha k^\beta\Big)
{\rm Tr}
\delta [\varep-\hat{\xi}( {\bf p})]
\hat{T}^R(\varep, {\bf p}, {\bf k})
\delta [\varep-\hat{\xi}( {\bf k})]
\hat{T}^A(\varep, {\bf k}, {\bf p}).
\label{a15a}
\eea
and the 
first term in
(\ref{a15a}) is precisely the contribution which describes the
effect of the gapless excitations -- orthogonality catastrophe. It is
present for the constant density of states.
For the spherically symmetric case, one can easily recover frome Eq.~(\ref{a15a})
 the
exponential factor in Eq.~(\ref{expansion.z}).

The second term in the expression for the Action characterizes the
electron-hole asymmetry. It has an explicit expression
\bea
&&Q^{\alpha\beta}
=\int \frac{d^3 p}{(2\pi)^3}
\frac{d^3 k}{(2\pi)^3}
\Big(p^\alpha p^\beta - p^\alpha k^\beta\Big)
\Big(Q_1 + Q_2\Big);
\label{a15b}
\\
&&Q_1= {\rm Im}
\int\frac{d\varep}{\pi} \tanh \frac{\varep}{2T}{\rm Tr}\
\frac{\p \hat{G}_0^R(\varep, {\bf p})}{\p \varep}
\hat{T}^R(\varep, {\bf p}, {\bf k})
\frac{\p\hat{G}_0^R(\varep, {\bf k})}{\p \varep}
\hat{T}^R(\varep, {\bf k}, {\bf p})
\nonumber\\
&&Q_2=  2\pi {\rm Re}\int{d\varep} \frac{d}{d\varep}\tanh
\frac{\varep}{2T}
{\rm Tr}\Bigg\{
\delta [\varep-\hat{\xi}( {\bf k})]\hat{T}^A(\varep, {\bf k}, {\bf p})
\Bigg[
\p_\varep \hat{T}^R(\varep, {\bf p}, {\bf k})
\delta [\varep-\hat{\xi}( {\bf p})]
+\frac{\Big[{\rm Re}\p_\varep \hat{G}_0^R(\varep, {\bf p})\Big]}{\pi}
\hat{T}^R(\varep, {\bf p}, {\bf k})
\Bigg]\Bigg\}
\nonumber
\eea
where the term $Q_1$ depends on the spectrum only
and $Q_2$ is due to the frequency dependence of the kinetic coefficients.
\end{mathletters}
Equation (\ref{a15b}) vanishes for the constant density of states
approximation, but it is not so for the arbitrary band structure.
However, we saw already that it generates the 
contribution proportional to the higher power of $\omega_n$. 
Thus, this term
produces {\em non-singular} correction to the leading term.
The characteristic value of this correction may be estimated as
$\simeq 1/(\tau_{tun}\epsilon^*)$, where $\epsilon^*$ is the energy
scale governing the electron-hole asymmetry. In principle, it might be
estimated from the thermopower measurements or from the first
principle band structure calculation of Eq.~(\ref{a15b}).

One can proceed with the similar expansion in the next orders of
perturbation theory to recover the constant $\Lambda_x$. 
The structure remains the
same, the first non-analytic contribution in $\omega_n$ comes from
terms  which are kept in the constant density of states approximation, and
the next contribution has higher power of $\omega_n$, and thus gives
the correction small as $ 1/(\tau_{tun}\epsilon^*)$. 
Thus, the electron-hole asymmetry (if treated systematically) can not
produce anything except parametrically small corrections to the
coupling constants 
in contradiction to claims of Ref.~\onlinecite{ZZ2}.

\begin{multicols}{2}

\end{multicols}


\begin{references}
\bibitem{footnote} The form of the regularization of the algebraically
divergent integrals in (\ref{zrho1}) and (\ref{xz3}) is unambiguously 
dictated by
the requirement that the result should vanish in the adiabatic limit
$\tau_{tun} \simeq 1/T$.
\bibitem{finiteT}
An extension of these results to finite temperature can be obtained via the 
conformal transformation 
\[
z\rightarrow z'=\frac{i}{2\pi T}\ln z
\]
with $z=\tau+ix$. This is analytic in the complex plane except the 
origin and maps the complex $z$-plane (without the origin) into the strip of 
width $1/T$ with periodic boundary conditions, which is the geometry 
that describes a 
system at finite temperature. Then the two point function 
in the strip can be obtained from the one in the plane using the 
transformation properties of primary fields under a conformal 
transformation (see for instance \cite{book.GNT}). For example, 
the correlation functions of vertex operators in a strip geometry, 
needed in (\ref{expansion}) at finite temperature, can be obtained
from the one on the whole complex plane via the substitution
\[
1/\tau \rightarrow \pi T/\sin\pi T\tau.
\]
Then  the partition function of the LGM at finite temperature
will have the form  (\ref{zfinalspin}-\ref{hspin}) with
\[
U_2(\tau) \rightarrow 
\int_0^{1/T} d\tau_1 \;
\eta_2(\tau_1)
\ln \Big|\frac{\sin \pi T (\tau_1-\tau)}{\pi T \tau_{tun}}\Big|.
\]
\end{references}
\end{document}